\documentstyle[12pt]{ioplppt}          
\begin{document}
\jl{4}
\title{Spectroscopic Amplitudes for One--Nucleon Transfer Between
1p0f--Shell Nuclei}[Spectroscopic Amplitudes for One--Nucleon Transfer Between
1p0f--Shell Nuclei]
\author{E. Kwa{\'s}niewicz\dag, H. Herndl\ddag~and H. Oberhummer\ddag}
\address{\dag\ Institute of Physics, Silesian Technical University,
Krasinskiego 8, PL--40 019 Katowice, Poland}
\address{\ddag\ Institute for Nuclear Physics, Technical University Vienna,
Wiedner Hauptstrasse 8--10, A--1040--Wien, Austria}
\begin{abstract}
Spectroscopic amplitudes are calculated for one--nucleon transfer between
low--lying, normal--parity states of nuclei in the lower part of the
1p$0\,$f--shell. Calculations are performed using shell--model wave
functions obtained from the diagonalization procedure of a nuclear
Hamiltonian in the space given by the complete set
of states generated from the 1p$_{3/2}$, 1p$_{1/2}$, $0\,$f$_{7/2}$ and
$0\,$f$_{5/2}$ orbits. The Hamiltonian contains
one and two body interactions derived recently by Richter \etal. Sum rules for
one--nucleon pick--up and stripping reactions are given. The selectivity in
excitation of the final
states induced by one--nucleon pick--up or stripping is discussed.
\end{abstract}
\maketitle
\section{Introduction}
Over the last thirty years spectroscopic amplitudes (SA's)
have been widely employed in many semi--microscopic studies on nuclear
structure and transfer reaction
mechanisms. These investigations have been performed in the framework
of the Distorted Wave
Born Approximation (DWBA), because of its advantage to relate in a simple
way the kinematic and
spectroscopic conditions of transfer reactions (Austern 1970, Satchler 1970,
Glendenning 1983).

On the other hand noticeable progress has been attained in the developement
of fully
microscopic methods to describe nuclear structure and reactions (Wildermuth
and Tang 1977, Hofmann 1987)
without the necessity to use spectroscopic amplitudes
at all. However, in spite of this fact, there are limitations in applying
fully microscopic
methods to systems with a small number of nucleons. Therefore, SA's are
still indispensable
in nuclear structure and reaction investigations.

In the literature many papers can be found which consider SA's
for nuclei from a wide mass range.
The most complete and consistent data exist for 0p--shell
nuclei (Cohen and Kurath 1967, Cohen and Kurath 1970,
Kurath and Millener 1975, Kurath 1973, Kwa{\'s}niewicz and Jarczyk 1985,
Kwa{\'s}niewicz \etal 1985).
Also a noticeable collection of multinucleon SA's is given for
1s0d--shell nuclei (Conze and Monakos 1979, Glaudemans \etal 1964,
McGrory and Wildenthal 1973, Inoue \etal 1966, Hecht and
Braunschweig 1975, Draayer 1975). On the
other hand, for heavier nuclei, the existing data are neither
complete nor consistent. In many cases calculations are limited to selected
nuclei with
wave functions created by imposing various limitations on the model space
and by employing various interactions (McGrory 1970, Meuders \etal 1976,
Koops and Glaudemans 1977, Van Hees and Glaudemans 1979, Kutschera \etal 1978,
Poves and Zuker 1981, Muto \etal 1978, Cole 1985, McCullen \etal 1964).
The necessity to employ various truncated model spaces is dictated by the
diagonalization
procedure of matrices with enormously large dimensions.

Recently new two--body interactions for 1p$0\,$f--shell nuclei have been
derived by Richter
\etal 1991. With these interactions the shell--model calculations have been
carried out
in the full shell--model space generated from the
1p$_{3/2}$, 1p$_{1/2}$, $0\,$f$_{7/2}$ and $0\,$f$_{5/2}$ orbits.
A satisfactory description of the experimental binding
energies, energy spectra, magnetic dipole and electric quadrupole moments
of nuclei
in the lower part of the 1p$0\,$f--shell is obtained.
This implies to use the wave functions yielded by the new interactions in
calculations
of other observable quantities.

The aim of this work is i) to create a consistent set of one--nucleon SA's
for low--lying
normal parity states of the 1p$0\,$f--shell nuclei (in the mass range
$41\leq A\leq $ 47)
by employing the wave functions obtained with the interaction of
Richter \etal 1991, ii) to show simple examples of applications of these
data in predicting the intensity in population of states of nuclei
which are excited in transfer reactions.
The restriction to nuclei from the lower part of 1p$0\,$f--shell is imposed by
long CPU--time and enormous disc quota required on the computer (CONVEX C3)
to solve the eigenvalue problem for heavier nuclei.

The organisation of the paper is as follows. In section 2 a method of
calculation
for one nucleon
SA's for 1p$0\,$f--shell nuclei is given.
In section 3 the sum rules for calculated SA's are formulated and selected
examples of their application to predict the population of the final states
produced in one nucleon pick--up
and stripping reactions are presented. The results are summarized in section 4.
\section{Formalism}
\subsection{Spectroscopic amplitudes}
The definition of SA's can be found in many papers. According to
the notations of Towner 1977 and Ichimura \etal 1973
the SA's for decomposition of a nucleus $A$
into the core $B$ and a nucleon can be written as\footnote
{The capital letters $A$ and $B$ denote the nucleus or its mass number
depending
on the context}
\begin{equation}
S_{n\ell j}^{\frac{1}{2}}=\sqrt{A}\,\langle\Phi_{E_A J_A T_A } |
\left(\Phi_{E_B J_B T_B }\times \phi_{n\ell j\tau}(\bf{r} )\right)^{J_A T_A}
\rangle \, ,\label{1}
\end{equation}
where $\Phi_{E_k J_k T_k}$ is the fully antisymmetric, intrinsic wave
function of a nucleus
$k$ ($k=A$ or $B$) labelled by the exitation energy $E_k$, spin $J_k$ and
isospin $T_k$
(third components of angular momenta are supressed) and $\phi_{n\ell
j\tau}(\bf{r} )$
is the wave function of a nucleon in the state of relative motion with respect
to the nucleus $B$ specified by the number of nodes $n$ and angular
momentum $l$.
The quantum numbers $j$ and $\tau$ define the spin (${\bf j} = {\bf \ell} +
{\bf s}$)
and charge of a nucleon. The coordinate $\bf{r}$ denotes the distance between
the center of
mass of nucleus $B$ and the separated nucleon.
Assuming that the intrinsic nuclear wave functions $\Phi_{E_k J_k T_k}$ are
approximated by the internal
part of the shell--model wave functions $\Psi_{E_k J_k T_k}$, equation
(\ref{1}) can be
transformed to the following expression (Towner 1977)
\begin{equation}
S_{n\ell
j}^{\frac{1}{2}}=\sqrt{A}\,\left(\frac{A}{A-1}\right)^\frac{2n+\ell }{2}
\,\langle\Psi_{E_A J_A T_A } |
\left(\Psi_{E_B J_B T_B }\times \phi_{n\ell j\tau}(\bf{r}_{\rm c})
\right)^{J_A T_A}
\rangle \, ,\label{2}
\end{equation}
where now the wave function $ \phi_{n\ell j\tau}(\bf{r}_{\rm c})$ depends
on the coordinate
$\bf{r}_{\rm c}$ of the separated nucleon in the laboratory frame of reference.

In this work the SA's are considered for normal parity states of nuclei
belonging
to the (1p$0\,$f$)^n$ shell--model space generated from the 1p$_{3/2}$,
1p$_{1/2}$,
$0\,$f$_{7/2}$ and $0\,$f$_{5/2}$
orbits. Thus the basis states can be written
\begin{equation}
|\rho\Gamma\rangle =|({\rm core})^{m_{\rm c}}(\sigma_1^{m^\rho_1})
(\sigma_2^{m^\rho_2})(\sigma_3^{m^\rho_3})
(\sigma_4^{m^\rho_4})\Gamma_1^\rho\Gamma_2^\rho\Gamma_3^\rho\Gamma_4^\rho
\Gamma_{12}^\rho\Gamma_{34}^\rho\, ;\Gamma\rangle\, ,\label{3}
\end{equation}
where $\rho$ is the running index used to number basis states. The
$\sigma_k$ $(k=1,2,3$ or 4)
represent a set of quantum numbers $n_k,\ell _k$ and $j_k$ to define subshells
1p$_{3/2}$,
1p$_{1/2}$, $0\,$f$_{7/2}$ and $0\,$f$_{5/2}$, respectively, and $m_k^\rho$
describes
the number of nucleons in the subshell $\sigma_k$ whereas $m_{\rm c}$ is
the number
of nucleons in the entirely filled up core. The $\Gamma^{\rho}_k$ gives a
set of spin and isospin quantum numbers (and additional
quantum numbers) describing a configuration of $m_k^\rho$ nucleons in the
subshell $\sigma_k$. The symbols
${\bf{\Gamma}}_{12}^\rho={\bf{\Gamma}}_1^\rho+{\bf{\Gamma}}_2^\rho$ and
${\bf{\Gamma}}_{34}^\rho={\bf{\Gamma}}_3^\rho+{\bf{\Gamma}}_4^\rho$  denote
the intermediate coupling
spin and isospin angular momenta quantum numbers. Finally ${\bf{
\Gamma}}={\bf{\Gamma}}_{12}+
{\bf{\Gamma}}_{34}$ defines the total spin and isospin of the state
$|\rho\Gamma\rangle$.
Expanding the wave function $\Psi_{E_k J_k T_k}$ in the basis (\ref{3})
the SA's of equation (\ref{2}) for a separation of one nucleon from the orbital
$\sigma_k$ have the form
\begin{equation}
S^{\frac{1}{2}}(\sigma_k)
=\sqrt{A}\left(\frac{A}{A-1}\right)^{\frac{2n_k+\ell _k}{2}}\sum C_{ A}
(i)C_{ B}(f)S_{if}^{\frac{1}{2}}(\sigma_k),\label{4}
\end{equation}
where $C_{ A}$ and $C_{ B}$ are the expansion coefficients of wave
functions of nuclei $A$ and $B$.
The overlap integral
\begin{eqnarray}
\fl S_{if}^{\frac{1}{2}}(\sigma_k) = \langle ({\rm core})^{m_{\rm c}}
(\sigma_1)^{m_1^i}(\sigma_2)^{m_2^i}(\sigma_3)^{m_3^i}(\sigma_4)^{m_4^i}
\Gamma_1^i\Gamma_2^i\Gamma_3^i\Gamma_4^i\Gamma_{12}^i\Gamma_{34}^i\Gamma_{ A}
\,|\nonumber\\
\lo{\times}|
[( ({\rm core})^{m_{\rm c}}
(\sigma_1)^{m_1^f}(\sigma_2)^{m_2^f}(\sigma_3)^{m_3^f}(\sigma_4)^{m_4^f}
\Gamma_1^f\Gamma_2^f\Gamma_3^f\Gamma_4^f\Gamma_{12}^f\Gamma_{34}^f\Gamma_{
B}\,)
\times (\sigma_k)]^{\Gamma_{\hat A}}\rangle\label{5}
\end{eqnarray}
describes a nucleon transition between basis states $|i\Gamma_{A}\rangle$ and
$|f\Gamma_{B}\rangle$. The explicit formulae for the $S_{if}(\sigma_k)$ are
given in the appendix.
\subsection{Sum rules}
The sum rules for one--nucleon SA's have first been considered by
Macfarlane and French 1960.
However, these sum rules have been derived only for particular cases of
nuclear states
expanded in the basis generated from one to two active orbits. In this work
nuclear
states are considered in a complete basis generated from four active orbits
in the 1p$0\,$f--shell and the sum rules for SA's calculated in the full
1p$0\,$f
shell--model space have to be considered.

The sum rule for one--nucleon pick--up from the target $A$ leading to states
of a nucleus
$B=A-$1 is given by
\begin{equation}
\sum_{ B}\;S_{ A\rightarrow B=A-1}(\sigma_k)=\sum_i C^2_{ A}(i)m_k^i
=\langle m_k\rangle, \label{6}
\end{equation}
where the summation on the left--hand side runs over all states of the
nucleus $B$
which give non--zero
SA's for one--nucleon pick--up from the orbit $\sigma_k$ in the nucleus $A$.
The expansion coefficients of the wave function of a nucleus $A$
in the ground state are denoted by $C_{ A}$.
The number of nucleons in the orbit $\sigma_k$ of the basis state
$|i\Gamma_A\rangle$
is given by $m_k^i$. The sum rule of equation (\ref{6}) gives an average number
$\langle m_k\rangle$
of nucleons in the orbit $\sigma_k$ which can take part in one--nucleon
pick--up from the
target nucleus $A$.
If additionally a summation over all active orbits $\sigma_k$ in both sides of
equation (\ref{6}) is performed one obtains
\begin{equation}
\sum_{{ B},\sigma_k}\, S_{ A\rightarrow B=A-1}(\sigma_k)=n\, ,\label{7}
\end{equation}
where $n$ is the total number of nucleons in all active orbits in the
nucleus $A$.

The sum rule (\ref{7}) provides a very simple check for calculations of SA's.
Furthermore a total strength (Kurath and Millener 1975, Kwa{\'s}niewicz and
Jarczyk 1985)
is defined, i.e.~a quantity which can
be useful in predicting the intensity of the population of states of
residual nuclei produced by one--nucleon pick--up.
This will be discussed in section 3.2.

The sum rule for one--nucleon stripping on a target $A$ leading to states of
a nucleus $B=A+$1 is given by
\begin{equation}
\sum_{ B}\;\left(\frac{\hat\Gamma_{ B}}{\hat\Gamma_{ A}}\right)^2
S_{ A\rightarrow B=A+1}(\sigma_k)\,=\,\sum_iC^2_{ A}(i)
\left(N(\sigma_k)-m_k^i \right),\label{8}
\end{equation}
where $\hat \Gamma_x=\sqrt{(2J_x+1)(2T_x+1)}$ ($x=A$ or $B$) and $N(\sigma_k)=
2\,(2j_k+1)$
with $j_k$ denoting the total spin of a nucleon in the orbit $\sigma_k$.
In a manner similar to equation (\ref{7}) we sum over all active orbits
$\sigma_k$ and obtain from equation (\ref{8})
\begin{equation}
\sum_{{ B},\sigma_k}\, \left(\frac{\hat\Gamma_{ B}}{\hat\Gamma_{ A}}\right)^2
S_{ A\rightarrow B=A+1}(\sigma_k)\,=\,\sum _{\sigma_k}
N(\sigma_k)-n+1,\label{9}
\end{equation}
where $n$ is the total number of nucleons in active orbits in the final nucleus
$B=A+1$. Similar to equation (\ref{7}) the generalized sum rule (9) provides an
additional check for calculations of SA's for one--nucleon stripping and can be
employed in predicting the population of states of final nucleus $B$  produced
by one--nucleon stripping on the target $A$ (see discussion in section 3.2.).

The sum rules of equations (\ref{6}--\ref{9}) have been obtained with the help
of formulae (4) and (A4--A7)
by employing the orthonormality conditions for the
U--coefficients
and for the coefficients of fractional parentage (c.f.p.)
$\langle(\sigma_k)^{m_k} \Gamma_k|\}(\sigma_k)^{m_k-1} \Gamma_k\rangle$.
In addition, a particle--hole relation for c.f.p. (MacFarlane
and French 1960), i.e.~the relation
\begin{eqnarray}
\fl \langle(\sigma_k)^{m_k}\Gamma_k^f|\}(\sigma_k)^{m_k-1} \Gamma_k^i\rangle=
\frac{(N(\sigma_k)-m_k+1)(\hat\Gamma_k^i)^2}{m_k(\hat\Gamma_k^f)^2}\nonumber\\
\lo\times \langle(\sigma_k)^{(N(\sigma_k)-m_k+1)}\Gamma_k^i|\}(\sigma_k)
^{N(\sigma_k)-m_k}
\Gamma_k^f\rangle,\label{10}
\end{eqnarray}
where index $i(f)$ corresponds to the nucleus $A\, (B=A+1)$, has been utilized
to derive the sum rules for stripping.
Furthermore the orthonormality and closure relations for the expansion
coefficients
of the nuclear wave functions have been exploited in deriving the sum rules
for pick--up and stripping.
The sum rules (\ref{6}--\ref{9}) relate to SA's of equation (\ref{4}) where the
$\left({ A}/{ A-1}\right)^{{(2n+\ell)}/{2}}$ factor is ignored.
\section{Results and discussion}
\subsection{Calculations}
The SA's for one--nucleon transfer have been calculated for nuclei in the
mass range
$41\leq { A}\leq 47$ according to equation (\ref{4}) and equations
(A4--A7). The eigenfunctions
of nuclear states considered have been obtained with the help
of the program RITSSCHIL (Zwarts 1985)
by diagonalization of a nuclear hamiltonian in the space defined
by equation (\ref{3}). The upper limit of the mass number is constrained
because of the long CPU time and enormous disc quota
required on the CONVEX C3 computer to calculate the eigenenergies and
eigenfunctions. The single particle
energies for 1p$_{3/2}$, 1p$_{1/2}$, $0\,$f$_{7/2}$ and $0\,$f$_{5/2}$ orbits
and two--body matrix
elements, i.e.~the FPD6 interactions derived by Richter \etal 1991 have
been employed
in the present calculations. The procedure for the mass--dependence correction
of the two body matrix elements (Richter \etal 1991)
has been taken into consideration. A selected example
of the spectroscopic amplitudes for one--nucleon stripping from
$^{45}$Sc($7/2^-3/2$) is shown in table 1.
We have also calculated SA's for one--nucleon transfer
between excited states of both colliding nuclei.
These data can be useful for studying more sophisticated
transfer processes, e.~g. processes which occur inside stars.
The SA's for one--nucleon pick--up and stripping for target nuclei being in
the ground and excited
states can be obtained upon request from one of the authors (H.H.).
\subsection{The intensity of states populated in one--nucleon transfer}
Transfer reactions show a remarkable degree of selectivity by occupying only
a few states in the residual nuclei with any appreciable strength. This
provides
an attractive way to relate the dynamic and spectroscopic conditions which
are clearly exhibited in the framework of the Distorted Wave Born Approximation
(DWBA) for direct nucleon transfer. The transfer amplitude in the simplest
case is
a product of the dynamic form factor calculated with
DWBA in the optical model and the SA (Austern 1970,
Satchler 1970, Glendenning 1983). Due to this factorization
one can determine the spectroscopic factor (i.e.~square
of the SA) as a quotient of the measured value of the cross
section $d\sigma / d\Omega$ and the theoretical value of
the squared DWBA dynamic form factor $\sigma^{\rm DWBA}(\Theta)$
(Glendenning 1983). The spectroscopic factor determined in
this way will be denoted as the experimental spectroscopic
factor. It can be compared with the theoretical spectroscopic
factor obtained from a shell--model description of the
initial and final nuclear states.

In this section we focus on an examination of the spectroscopic conditions
which are reflected in the nuclear overlap integrals considered in section 2.
Neglecting the dynamic conditions
the transition
intensity can be deduced from considering the percentage distribution of the
total strength (defined for pick--up by the sum rule of equation (\ref{7})
and for stripping by the sum rule of equation (\ref{9}))
among states populated by one--nucleon transfer.

Figure 1 illustrates an example where almost all total strength is
deposited in the lowest
states
of the residual nucleus. In a nucleon pick--up
from $^{43}$Ca 94\% of the total strength is deposited in the six lowest
lying states
of $^{42}$Ca from among which the first
$J^\pi T=0^+1$ and the first $6^+1$ states  absorb 25 \%
and 33\% of the total strength respectively. A striking example is
neutron pick--up from the $^{44}$Ca target. In this case the calculation of the
SA's gives that 91\% of the total strength is deposited in
the first 7/2$^-3/2$ state of the residual nucleus $^{43}$Ca.

Examples where around half of the total strength is deposited in the first few
states of the residual
nucleus are presented in figures 2 and 3. The neutron pick--up from the
$^{45}$Sc
target gives 50\% of the total strength distributed among the first
2$^+1$ (7\%),
6$^+1$ (8\%), 4$^+1$ (7\%), 1$^+1$ (6\%), 7$^+1$ (22\%)
states of the final $^{44}$Sc nucleus (figure 2).

Theoretical predictions are in excellent agreement with experimental findings
(see table 2) obtained from studying the (d,t) reaction on
the $^{45}$Sc target (Ohnuma and Sourkes 1971).
Here we want to mention that the spectroscopic factor for the
isobaric analog of the $^{44}$Ca$_{\rm g.s.}$ in $^{44}$Sc
($J^{\Pi}T$, $E_{\rm exp} = 0^{+}2$, 2.783\,MeV) extracted from
$^{45}$Sc($^{3}$He,$\alpha$)$^{44}$Sc (Rappaport \etal 1971)
and $^{45}$Sc(d,t)$^{44}$Sc (Ohnuma and Sourkes 1971)
is equal to 0.5 and 1.1, respectively, while from our calculations
a value of 0.55 is obtained. In the $0^{+}2$ state excited
by neutron pick--up from the $^{45}$Sc target 11\% of the total
stength is deposited. In the first excited state of $^{45}$Sc
($J^{\Pi}T$, $E_{\rm theor} = 2^{+}2$, 4.755\,MeV)
only 1\% and in the next 10 excited $T=2$ states
our shell--model calculations predict altogether
less than 1\% of the total strength.
About a quarter of the total strength is distributed among the seven
lowest states of $^{47}$Ca
produced by the neutron stripping on the $^{46}$Ca target (figure 3).
A total strength of 23\% is deposited in the first
7/2$^-$7/2 (7\%),
the first 3/2$^-$7/2 (11\%)
and the second 1/2$^-$7/2 (5\%) states of $^{47}$Ca.

An example of relatively small concentration of the total strength in the
lowest
lying states of
the residual nucleus is illustrated in figure 4. Namely, considering the
neutron
stripping on the
$^{45}$Sc target, only 12.9\% of the total strength is distributed among
the seven
lowest states of the $^{46}$Sc nucleus.
The experimental transition strength $(2J_f+1)S/(2J_i+1)$ for above mentioned
states of
$^{46}$Sc which are induced by the (d,p) (Roy \etal 1992,
Rappaport \etal 1966, Bing \etal 1976) and (t,d) (Brussaard and Glaudemans
1977)
reactions on the $^{45}$Sc target is compared with the theoretical predictions
(figure 4 and table 3).
Theoretical results of the present work are also compared with the earlier
results obtained by employing the truncated shell--model space generated
from the $0\,$f$_{7/2}$ orbit only (McCullen \etal 1964).
Because of experimental uncertainities (Roy \etal 1992) it is difficult to
determine which
theoretical calculations better describe the experimental data. It seems that
both theoretical calculations (figure 4) are within limits of
uncertainity in good agreement
with the experimental results.

In order to get a better insight into the quality of the present calculations,
the theoretical predictions of spectroscopic factors of
one--nucleon pick--up and stripping are compared with the
experimental findings in tables 2 and 3. As already mentioned, the experimental
spectroscopic factors were obtained under the assumption that only one
orbital $(nlj)$ contributes to the cross section. This assumption
seems to be justified since the magnitudes of the theoretical SA's
assigned to this orbital considerably exceeds the ones
assigned to the remaining allowed orbitals.
Therefore, for instance, in the $^{45}$Sc(d,t)$^{44}$Sc
and $^{45}$Sc(d,p)$^{46}$Sc reactions almost only the
$nlj=0\,3\,7/2$ orbital is populated.

Spectroscopic factors deduced from one--nucleon transfer reactions
are as a rule sensitive to the radius of the potential in which the
bound--state wave function is generated as well as to the
optical model potential describing wave functions in the entrance
and exit reaction channel.

During the last decade many nuclei have been studies with
electron--induced proton knock--out (e,e'p)--reactions (de Witt Huberts 1990).
An important advantage is that the spectroscopic factors
deduced from these reactions are not sensitive to the shape
of the single--particle binding potential and to the optical potential
describing the scattering state of the outgoing proton. However,
the spectroscopic factors determined from (e,e'p)--reactions are
surprisingly small compared to results obtained with transfer
reactions, sometimes even by 30--40\% (de Witt Huberts 1990).
Theoretical investigations initiated to understand this effect
(Pandharipande 1989, Benhar \etal 1989, Van Neck \etal 1991,
Wesseling \etal 1992) indicate that short--range correlations
and uncertainties of the (e,e'p)--reaction
mechanism, in particular due to the final state interaction sector
may be responsible for this effect (de Witt Huberts 1990). The lack of
reliability of the approximations inherent in
these models, in particular for light nuclei and medium--mass
systems, still does not seem to allow to make
significant conclusions on possible
differences of spectroscopic factors deduced from nuclear transfer
and (e,e'p)--reactions (for a more detailed discussion see
de Witt Huberts 1990). Anyway, the (e,e'p)--reactions as
a spectroscopic tool exhibit more model sensitivity
than transfer reactions when studying phenomena of
multinucleon systems .
\subsection{Fluctuations in spectroscopic factors}
Some applications of the sum rules for SA's allow obtaining
valuable informations on nuclear structure without performing
advanced shell--model calculations. As an illustrative example
can serve the case when
the summations in the sum rules (\ref{7}) and (\ref{9}) reduce to a single
term yielding expressions for individual spectroscopic
factors. Due to this property fluctuations in spectroscopic
factors throughout a shell/subshell can be demonstrated.

Consider $m$ identical nucleons occupying a single shell
$\sigma_{k}$ above the $^{40}$Ca core. Applying equation
(\ref{9}) to the
$\left(\sigma_{k}\right)^{2m_k+1}_{J=j_k} \rightarrow
\left(\sigma_{k}\right)^{2m_k}_{J=0}$ transition one obtains
\begin{equation}
(2j_k+1)S\left(2m_k+1 \rightarrow 2m_k\right)=
\left(2m_k+1\right)-2m_k. \label{11}
\end{equation}
Similarly equation (\ref{7}) yields for the
$\left(\sigma_{k}\right)^{2m_k}_{J=0} \rightarrow
\left(\sigma_{k}\right)^{2m_k-1}_{J=j_k}$ transition
\begin{equation}
S\left(2m_k \rightarrow 2m_k-1\right)=2m_k. \label{12}
\end{equation}
Thus for the chain of transitions
$$\left(m_{k} J\right) = (00) \leftarrow \left(1j_k\right)
\leftarrow (20) \leftarrow \left(3j_k\right) \ldots$$
expressions (\ref{11}) and (\ref{12}) combine to give
\begin{equation}
\label{cases}
S\left(m_k \rightarrow m_k-1\right)=
\cases{1-\frac{m_k-1}{2j_k+1} & for $m_k$ odd,\\
m_k & for $m_k$ even.\\} \label{13}
\end{equation}

With the aid of equations (\ref{13}) the chain of spectroscopic
factors for
$\left(\sigma_k\right)^{m_k} =
\left(0{\rm f}_{7/2}\right)^{m_k}$ configurations concerning
the ground states in Ca can immediately
be calculated (table 4). A comparison of these spectroscopic
factors with their equivalents calculated in the full
$\left(1{\rm p}0{\rm f}\right)^m$ shell--model space (table 4)
allows to conclude that the structure
of the ground states of $^{41}$Ca--$^{45}$Ca nuclei is dominated
by the $\left(0{\rm f}_{7/2}\right)^{m}$ configuration. This is
contrary to the structure of the ground states of $^{46}$Ca
and $^{47}$Ca nuclei which differs considerably from a simple
$\left(0{\rm f}_{7/2}\right)^{m}$ configuration.
\section{Summary}
Spectroscopic amplitudes for one--nucleon pick--up and one--nucleon
stripping reactions with nuclei from the lower part of 1p$0\,$f--shell
are calculated
with the help of wave functions expanded in the complete basis generated
from the
1p$_{3/2}$, 1p$_{1/2}$, $0\,$f$_{7/2}$ and $0\,$f$_{5/2}$ orbits.
The two--body mass--dependent FPD interactions recently derived by Richter
\etal 1991
have been adapted in the calculations of the wave functions. The sum rules
for calculated SA's have been derived and examples of their application in
predicting the
intensity of states of the final nucleus produced by one--nucleon pick--up
and stripping reactions
are discussed. The examples indicate that for some reactions nearly 100\%
of the total strength
is concentrated in a few low--lying states of the residual nucleus (see
figure 1).
However, there are examples where the total strength is distributed over a
wider energy
range of excited states of the final nucleus.

The presented example indicate that the distribution of the total strength
among residual 1p$0\,$f--shell
nuclei is similar to that for the lighter nuclei (see for example
Cohen and Kurath 1967, Cohen and Kurath 1970,
Kurath and Millener 1975, Kurath 1973, Kwa{\'s}niewicz and Jarczyk 1985,
Kwa{\'s}niewicz \etal 1985).

\ack Partially supported by BK/BW/PSl--IF(Poland)
and by the Fonds zur F\"orderung
der wissenschaftlichen Forschung in \"Osterreich, project P8806--PHY.

-------------------------------------------------------------------------
| Heinz Oberhummer              | internet: hoberhum@ecxph.tuwien.ac.at  |
| Institut fuer Kernphysik      | tel:      +43/1/58801-5574 office      |

| Technische Universitaet Wien  |           +43/1/58801-5575 secretary   |
| Wiedner Hauptstrasse 8-10     | Fax number (till 15.10.94) +43/1/564203|

| A-1040 WIEN                   | NEW fax number: 0043/1/5864203         |
| Austria                       | Ttx:      (61)3222467=tuw              |
-------------------------------------------------------------------------

\begin{appendix}
\section{}
In this subsection a method of deriving the formulae for the partial
spectroscopic amplitudes
$S_{if}^\frac{1}{2}(\sigma_k)$ of equation (\ref{5})
is sketched. This is demonstrated for the simplest case, i.e.~when a
nucleon is transferred from
the last subshell $\sigma_4$. Introducing the diagrammatic notation of
Macfarlane and French 1960
the overlap integral $S_{if}^\frac{1}{2}(\sigma_4)$ of equation (5) is
defined below:
\begin{equation}
\vbox{
%
%
\vspace{5cm}}\label{A1}\end{equation}

In order to single out the groups of nucleons $(\sigma_k)^{m_k} (k$=1,2,3 or 4)
on both sides of (\ref{A1}), wave functions have to be constructed that are
antisymmetric in these groups
separately.
Following the procedure described in Brussaard and Glaudemans 1977 one obtains
\begin{equation}
\vbox{
%
%
\vspace{5cm}}\label{A2}\end{equation}
Employing the one--nucleon expansion for the ($\sigma_4)^{m_4^i}$ and
performing a change of angular
momenta coupling order the bra wave function of equation (\ref{A2}) can be
written
\begin{equation}
\vbox{
\vspace{12cm}}\label{A3}\end{equation}
%
%
{}From equations (\ref{A2}) and (\ref{A3}) one immediately obtains
\begin{eqnarray}
\fl S_{if}^{\frac{1}{2}}(\sigma_4) =   \left(\frac{m^i_4}{
A}\right)^\frac{1}{2}
\langle(\sigma_4)^{m^i_4}\Gamma_4^i|\}((\sigma_4)^{m_4^i-1}\Gamma_4^f\rangle
\nonumber\\
\lo
\times U(\Gamma_3^i\Gamma_4^f\Gamma_{34}^i\sigma_4;\Gamma_{34}^f\Gamma_4^i)\,
U(\Gamma_{12}^i\Gamma_{34}^f\Gamma_{ A}\sigma_4;\Gamma_{ B}\Gamma
_{34}^i)\nonumber\\
\lo
\times \delta_{\Gamma_1^i \Gamma_1^f}\delta_{\Gamma_2^i \Gamma_2^f}
\delta_{\Gamma_{12}^i \Gamma_{12}^f}\delta_{\Gamma_3^i \Gamma_3^f}
\delta_{m_1^i m_1^f}\delta_{m_2^i m_2^f}
\delta_{m_3^i m_3^f}\delta_{m_4^i-1 m_4^f}.
\end{eqnarray}
In the same way as above the following expressions are derived
\begin{eqnarray}
\fl S^\frac{1}{2}_{if}(\sigma_3)   =   \left(\frac{m_3^i}{
A}\right)^\frac{1}{2}
(-)^{\Gamma_3^i+\Gamma_3^f-\Gamma_{34}^i-\Gamma_{34}^f-m_4^i}
\langle(\sigma_3)^{m_3^i}\Gamma_3^i|\} (\sigma_3)^{m_3^i-1}\Gamma_3^{f}
\rangle\nonumber\\
\lo
\times U\left(\Gamma_4^i \Gamma_3^f\Gamma_{34}^i
\sigma_3;\Gamma_{34}^f \Gamma_{3}^i\right)\, U\left(\Gamma_{12}^i\Gamma_{34}^f
\Gamma_{ A}
\sigma_3;\Gamma_{ B}\Gamma_{34}^i\right)\nonumber\\
\lo
\times \delta_{\Gamma_1^i \Gamma_1^f}\delta_{\Gamma_2^i \Gamma_2^f}
\delta_{\Gamma_{12}^i \Gamma_{12}^f}\delta_{\Gamma_4^i \Gamma_4^f}
\delta_{m_1^i m_1^f}\delta_{m_2^i m_2^f}
\delta_{m_3^i-1, m_3^f}\delta_{m_4^i m_4^f}
\end{eqnarray}
\begin{eqnarray}
\fl S^\frac{1}{2}_{if}(\sigma_2)   =   \left(\frac{m_2^i}{
A}\right)^\frac{1}{2}
(-)^{m_3^i+m_4^i+\Gamma_{12}^i+\Gamma_{12}^f-\Gamma_{ A}-\Gamma_{ B}}
\langle(\sigma_2)^{m_2^i}\Gamma_2^i|\} (\sigma_2)^{m_2^i-1}\Gamma_2^{f}
\rangle\nonumber\\
\lo\times
U\left(\Gamma_1^i\Gamma_2^f\Gamma_{12}^i\sigma_2;
\Gamma_{12}^f\Gamma_2^i\right)\, U\left(\Gamma_{34}^i\Gamma_{12}^f\Gamma_{ A}
\sigma_2;\Gamma_{ B}\Gamma_{12}^i
\right)\nonumber\\
\lo\times
\delta_{\Gamma_1^i \Gamma_1^f}\delta_{\Gamma_3^i \Gamma_3^f}
\delta_{\Gamma_4^i \Gamma_4^f} \delta_{\Gamma_{34}^i \Gamma_{34}^f}
\delta_{m_1^i m_1^f}\delta_{m_2^i-1, m_2^f}
\delta_{m_3^i m_3^f}\delta_{m_4^i m_4^f}
\end{eqnarray}
and
\begin{eqnarray}
\fl S^\frac{1}{2}_{if}(\sigma_1)   =   \left(\frac{m_1^i}{
A}\right)^\frac{1}{2}
(-)^{m_2^i+m_3^i+m_4^i+\Gamma_{1}^i+\Gamma_{1}^f-\Gamma_{ A}-
\Gamma_{ B}}
\langle(\sigma_1)^{m_1^i}\Gamma_1^i|\} (\sigma_1)^{m_1^i-1}\Gamma_1^{f}
\rangle\nonumber\\
\lo
\times   U\left(\Gamma_2^i\Gamma_1^f\Gamma_{12}^i
\sigma_1;\Gamma_{12}^f\Gamma_1^i\right)\, U\left(\Gamma_{34}^i\Gamma_{12}^f
\Gamma_{ A}\sigma_1;\Gamma_{ B}\Gamma_{12}^i
\right)\nonumber\\
\lo
\times \delta_{\Gamma_3^i \Gamma_3^f}\delta_{\Gamma_4^i \Gamma_4^f}
\delta_{\Gamma_{34}^i \Gamma_{34}^f}\delta_{\Gamma_2^i \Gamma_2^f}
\delta_{m_1^i-1, m_1^f}\delta_{m_2^i m_2^f}
\delta_{m_3^i m_3^f}\delta_{m_4^i m_4^f}\, .
\end{eqnarray}
\end{appendix}
\References
\item[] Austern N 1970 {\it
Direct Nuclear Reaction Theories} (Whiley: New York)
\item[] Benhar O, Fabrocini A and Fantoni S 1989 \NP {\bf A505} 267
\item[] Bing O, Bonneaud G Magnac--Valtett D and Gerardin C 1976
\NP {\bf A257} 460
\item[] Brown G Denning A and Haigh J G B 1974
\NP {\bf A255} 267
\item[] Brussaard  P J and Glaudemans P W M 1977
{\it Shell Model Applications in Nuclear Spectroscopy}
(North Holland: Amsterdam)
\item[] Cohen S and Kurath D 1967 \NP {\bf A101} 1
\item[] Cohen S and Kurath D 1970 \NP {\bf A141} 145
\item[] Cole B J 1985 \JPG {\bf 11} 961
\item[] Conze M and Monakos P 1975 \JPG {\bf 5} 671
\item[] De Witt Huberts P K A 1990 \JPG {\bf 16} 507
\item[] Draayer J P 1975 \NP {\bf A237} 157
\item[] Glaudemans P W M, Wiechers G and Brussaard 1964
\NP {\bf 56} 548
\item[] Glendenning N K 1983 {\it
Direct Nuclear Reaction Theories} (Academic Press: New York)
\item[] Hecht K T and Braunschweig D 1975 \NP {\bf A244} 365
\item[] Hofmann H M, {\it Lecture Notes in Physics} 1987 {\bf 273} 243
\item[] Ichimura M Arima A Halbert E C and Teresawa T 1973
\NP {\bf A244} 365
\item[] Inoue T, Sebe T, Hagiwara H and Arima A 1966
\NP {\bf 85} 184
\item[] Koops J E and Glaudemans P W M 1977
\ZP {\bf A280} 181
\item[] Kurath D 1973 \PR {\bf C7} 1390
\item[] Kurath D and Millener D J 1975 \NP {\bf A238} 269
\item[] Kutschera W, Brown B A and Ogawa K 1978
{\it Revista Nuovo Cimento} {\bf 1} 1
\item[] Kwa{\'s}niewicz E and Jarczyk L 1985 \NP {\bf A441} 77
\item[] Kwa{\'s}niewicz E, Kisiel J and Jarczyk L 1985 {\it Acta Physica
Polonica}
{\bf B16} 947
\item[] Macfarlane M H and French J B 1960 \RMP {\bf 32} 567
\item[] McCullen J D Bayman B F and Zamick L 1964
\PR {\bf B134} 515
\item[] McGrory J B and Wildenthal B H 1973 \PR {\bf C7} 974
\item[] McGrory J B, Wildenthal B H and Halbert E C 1970
\PR {\bf C2} 186
\item[] Meuders F, Glaudemans P W M, van Hienen J F A
and Timmer G A 1976 \ZP {\bf A276} 113
\item[] Muto K, Oda T and Hovie H 1978
Progress Theoretical Physics {\bf 60} 1350
\item[] Ohnuma H and Sourkes A M 1971 \PR {\bf C3} 158
\item[] Pandharipande V R 1989 \NP {\bf A497} 43c
\item[] Poves A and Zuker A 1981
{\it Physics Reports} {\bf 70} 235
\item[] Rappaport J, Sperduto A and Buechner W 1966
\PR {\bf 151} 939
\item[] Rappaport J, Belote T A, Bainum D E and
Dorenbusch W E 1971
\NP {\bf A168} 177
\item[] Richter W A, Van Der Merwe M G Julies R E and Brown B A 1991
\NP {\bf A523} 325
\item[] Roy J N Majumder A R and Sen Gupta H M 1992
\PR {\bf C46} (1992) 144
\item[] Satchler G R 1983 {\it Direct Nuclear Reactions} (Clarendon: Oxford)
\item[] Towner I S 1977{\it A Shell Model Description of Light Nuclei}
(Clarendon: Oxford)
\item[] Van Hees A G M and Glaudemans P W M 1979
\ZP {\bf A293} 327
\item[] Van Neck D, Waroquier M and Ryckebusch J 1991 \NP {\bf A530} 347
\item[] Wesseling J, de Jager C W, Lapikas L, de Vries H,
Harakeh M N, Kalantar--Nayestanaki N, Fagg L W, Lindgren R A
and Van Neck D 1992 \NP {\bf A547} 519
\item[] Wildermuth K and Tang Y C 1977 {\it
A Unified Theory of the Nucleus} (Vieweg: Braunschweig) (and references
therein)
\item[] Zwarts D 1985 {\it Computer Physics Communications} {\bf 38} 365
\endrefs
\Tables
\begin{table}
\caption[Table 1]{Spectroscopic amplitudes for one--nucleon stripping from
$^{45}$Sc($7/2^-3/2$). States of the final nucleus are specified
by the spin, parity and isospin in the first
column and calculated excitation energy in the second column. The
orbitals of the transferred nucleon are given in the next four columns.
In order to be in an agreement with the definition of equation (1)
all listed spectroscopic amplitudes have to be multiplied by the
$\left(A/(A-1)\right)^3/2$ factor.}
\lineup
\begin{indented}
\item[]
\begin{tabular}{@{}lllllll}
\br
$J^{\Pi}T$&$E_{\rm calc}$ &$P_{1/2}$ &$P_{3/2}$ &F$_{5/2}$ &F$_{7/2}$
&\hbox{percentage}\\
&(\hbox{MeV})&&&&&\hbox{of strength}\\
\mr
1$^+2$&1.788&       &       &0.169  &\-0.294 &0.15\\
1$^+2$&2.521&       &       &0.210  &0.131   &$<0.1$\\
2$^+2$&0.228&       &0.099  &\-0.122&0.554   &0.7\\
2$^+2$&1.679&       &0.116  &\-0.022&\-0.063 &$<0.1$\\
3$^+2$&0.203&0.042  &\-0.118&0.061  &\-0.703 &1.6\\
3$^+2$&1.278&\-0.021&\-0.031&0.041  &0.226   &0.17\\
4$^+2$&0.0  &0.024  &\-0.065&0.087  &\-0.598 &1.5\\
4$^+2$&0.784&0.036  &0.084  &\-0.004&0.489   &1.0\\
5$^+2$&0.331&       &\-0.185&\-0.048&\-0.600 &1.9\\
5$^+2$&0.894&       &\-0.057&0.053  &\-0.490 &1.2\\
6$^+2$&0.019&       &       &0.026  &\-0.866 &4.4\\
7$^+2$&0.892&       &       &       &0.449   &1.4\\
\br
\end{tabular}
\end{indented}
\end{table}
\begin{table}
\caption[Table 2]{A comparison of the experimental and calculated
spectroscopic factors for
selected one--nucleon pick--up reactions.
States of the final nucleus are specified by the experimental
excitation energy in the second column and spin and parity
in the third column. Orbitals of the transferred nucleon are given
in the fourth column.}
\lineup
\begin{indented}
\item[]
\begin{tabular}{@{}llllll}
\br
final nucleus&E$_x$(MeV)&$J^\Pi$&$\ell $j& $S$(exp.)& $S$(theor.)\\
\mr
$^{41}$Ca&0.0&$7/2^-$& 3 7/2&1.46$^{\hbox{(i)}}$ (1.60 $^{\hbox{(ii)}}$)&1.81\\
&1.942&$3/2^-$& 1 3/2&0.15$^{\hbox{(ii)}}$&0.11\\
$^{43}$Ca&0.0& $7/2^-$& 3 7/2&3.3$^{\hbox{(iii)}}$ (3.5
$^{\hbox{(iv)}}$)&3.64\\
&0.373&$ 5/2 ^-$&3 5/2&0.14$^{\hbox{(iii)}}$ (0.27 $^{\hbox{(iv)}}$)&0.03\\
&0.593&$ 3/2 ^-$&1 3/2 &0.09$^{\hbox{(iii)}}$ (0.14 $^{\hbox{(iv)}}$)&0.02\\
&2.046&$ 3/2 ^-$&1 3/2 &0.14$^{\hbox{(iii)}}$ (0.31 $^{\hbox{(iv)}}$)&0.14\\
$^{44}$Sc&0.0&$2^+$& 3 7/2&0.35$^{\hbox{(v)}}$&0.35\\
&0.271&$ 6^+$& 3 7/2&0.48$^{\hbox{(v)}}$&0.39\\
& 0.350&$ 4^+$& 3 7/2&0.35$^{\hbox{(v)}}$&0.34\\
&0.667&$ 1^+$& 3 7/2&0.32$^{\hbox{(v)}}$&0.29\\
&0.763&$ (3^+)$& 3 7/2&0.20$^{\hbox{(v)}}$&0.03\\
&0.968&$ 7^+$& 3 7/2&1.29$^{\hbox{(v)}}$&1.09\\
&1.052&$ (5^+)$& 3 7/2&0.25$^{\hbox{(v)}}$&0.14\\
&2.783&$ 0^+\,T=2$& 3 7/2&1.1$^{\hbox{(v)}}$ (0.5$^{\hbox{(vi)}})$&0.55\\
\br
\end{tabular}\\
$^{\hbox{(i)}}$ The $^{42}$Ca(p,d)$^{41}$Ca reaction ($E_{\rm p}$=26.5 MeV)
from Brown \etal 1974\\
$^{\hbox{(ii)}}$ The $^{42}$Ca($^{3}$He,$\alpha$)$^{41}$Ca reaction
$(E_{^{3}{\rm He}}$=18 MeV)
from Brown \etal 1974\\
$^{\hbox{(iii)}}$ The $^{44}$Ca(p,d)$^{43}$Ca reaction ($E_{\rm p}$=26.5 MeV)
from Brown \etal 1974\\
$^{\hbox{(iv)}}$ The $^{44}$Ca($^{3}$He,$\alpha)^{43}$Ca reaction
($E_{^{3}{\rm He}}$
=18 MeV) from Brown \etal 1974\\
$^{\hbox{(v)}}$ The $^{45}$Sc(d,t)$^{44}$Sc reaction ($E_{\rm p}$=19.5 MeV)
from Ohnuma and Sourkes 1971\\
$^{\hbox{(vi)}}$ The $^{45}$Sc($^{3}$He,$\alpha$)$^{44}$Sc reaction
from Rappaport \etal 1971
\end{indented}
\end{table}
\begin{table}
\caption[Table 3]{A comparison of the experimental and theoretical transfer
strengths
$(2J_f+1)S$ for selected
one--nucleon stripping reactions.
For explanation see caption to table 2.}
\lineup
\footnotesize\rm
\begin{tabular}{@{}llllll}
\br
final nucleus&E$_x$(MeV)&$J^\Pi$&$\ell $j& $(2J_f+1)S$(exp.)&
$(2J_f+1)S$(theor.)\\
\mr
$^{43}$Ca&0.0&$7/2^-$& 3 7/2&5.5$^{\hbox{(i)}}$&6.0\\
&0.593&$3/2^-$& 1  3/2&0.21$^{\hbox{(i)}}$&0.05\\
$^{45}$Ca&0.0&$7/2^-$& 3  7/2&3.40$^{\hbox{(ii)}}$&4.03\\
&1.435&$ 3/2^-$& 1  3/2&0.47$^{\hbox{(ii)}}$&0.13\\
&1.90&$ 3/2^-$& 1  3/2& 2.60$^{\hbox{(ii)}}$&3.39\\
&2.249&$ 1/2^-$& 1 1/2&0.36$^{\hbox{(ii)}}$& 0.24\\
$^{46}$Sc&0.0&$4^+$& 3  7/2&2.82$^{\hbox{(iii)}}$ (4.64
$^{\hbox{(iv)}}$)&3.22\\
&0.052&$ 6^+$& 3  7/2&6.90$^{\hbox{(iii)}}$ (10.64 $^{\hbox{(iv)}}$)&9.75\\
&0.228&$ 3^+$& 3  7/2&2.86$^{\hbox{(iii)}}$ (4.96 $^{\hbox{(iv)}}$)&3.46\\
&0.281&$ 5^+$& 1 3/2, 3 7/2&0.44, 4.29$^{\hbox{(iii)}}$ (0.8
$^{\hbox{(iv)}}$)&0.38, 3.96\\
&0.444&$ 2^+$& 3   7/2&1.60$^{\hbox{(iii)}}$ (2.48 $^{\hbox{(iv)}}$)&1.53\\
&0.774&$ 5^+$& 3  7/2&2.78$^{\hbox{(iii)}}$ (4.88 $^{\hbox{(iv)}}$)&2.64\\
&0.835&$ 4^+$& 3  7/2&1.68$^{\hbox{(iii)}}$ (2.56 $^{\hbox{(iv)}}$)&2.15\\
&0.977&$ 7^+$& 3  7/2&3.39$^{\hbox{(iii)}}$ (4.00 $^{\hbox{(iv)}}$)&3.02\\
\br
\end{tabular}\\
$^{\hbox{(i)}}$ The $^{42}$Ca(d,p)$^{43}$Ca reaction ($E_{\rm d}$=7 MeV) from
Brown \etal 1974\\
$^{\hbox{(ii)}}$ The $^{44}$Ca(d,p)$^{45}$Ca reaction ($E_{\rm d}$=7 MeV)
from Brown \etal 1974\\
$^{\hbox{(iii)}}$ The $^{45}$Sc(d,p)$^{46}$Sc reaction ($E_{\rm d}$=12 MeV)
from
ref. Roy \etal 1992\\
$^{\hbox{(iv)}}$ The $^{45}$Sc(d,p)$^{46}$Sc reaction ($E_{\rm d}$=7 MeV) from
Rappaport \etal 1966\\
\end{table}
\begin{table}
\caption[Table 4]{Comparison of the spectroscopic factors for
reactions involving ground states of Ca nuclei calculated:\\
(i) by assuming a pure $\left(0{\rm f}_{7/2}\right)^m$ configuration
for neutrons above the $^{40}$Ca core (second column),\\
(ii) in the full (1p0f)$^m$ shell--model space (third column).}
\lineup
\begin{indented}
\item[]
\begin{tabular}{@{}lll}
\br
&\centre{2}{spectroscopic factor}\\
Reaction&\crule{2}\\
& pure $\left(0{\rm f}_{7/2}\right)^m$ model &
full (1p0f)$^m$ model\\
\mr
$^{40}$Ca(d,p)$^{41}$Ca & 1 & 1\\
$^{42}$Ca(d,p)$^{43}$Ca & 0.75 & 0.75\\
$^{44}$Ca(d,p)$^{45}$Ca & 0.5 & 0.507\\
$^{46}$Ca(d,p)$^{47}$Ca & 0.25 & 1.036\\
$^{44}$Ca(p,d)$^{43}$Ca & 4 & 3.644\\
$^{42}$Ca(p,d)$^{41}$Ca & 2 & 1.812\\
\br
\end{tabular}\\
\end{indented}
\end{table}
\clearpage
\Figures
\begin{figure}
\caption[Figure 1]{The percentage distribution of the total strength among
states of
$^{42}$Ca produced by the neutron pick--up from the
$^{43}$Ca target. States of $^{42}$Ca are specified by the spin $J$,
parity $\Pi$, isospin $T$ and calculated excitation energy $E_{\rm calc}$.}
\end{figure}
\begin{figure}
\caption[Figure 2]{The percentage distribution of the total strength among
states of
$^{44}$Sc produced by the neutron pick--up from the $^{45}$Sc
target. Theoretical results (first columns) are compared with the
experimental data
(second ones) obtained from the $^{45}$Sc(d,t)$^{44}$Sc reaction
(Ohnuma and Sourkes 1971).}
\end{figure}
\begin{figure}
\caption[Figure 3]{The percentage distribution of the total strength among
states of
$^{47}$Ca produced by the neutron stripping on the
$^{46}$Ca target.}
\end{figure}
\begin{figure}
\caption[Figure 4]{The measured transition strength
$\left((2J_f+1)/(2J_i+1)\right)S$
for states of $^{46}$Sc produced by the neutron stripping on the $^{45}$Sc
target compared to the theory.\\
(a): theoretical predictions
from McCullen \etal 1964,\\
(b) (d,p) reaction from Roy \etal 1992,\\
(c) (d,p) reaction from Rappaport \etal 1966,\\
(d) (t,d) reaction from Bing \etal 1976,\\
(e) theory of the present work.}
\end{figure}
\end{document}